\documentclass[11pt,reqno,oneside]{article}
\usepackage{pstricks,pst-node,pst-tree}
\usepackage{amsmath}
\usepackage{amssymb}
\usepackage[lmargin=.55in, rmargin=.55in, tmargin=1in, bmargin=1in]{geometry}
\usepackage{xcolor}
\usepackage[dvips]{graphicx}

\usepackage{natbib}
\usepackage{pstricks}

\newtheorem{prop}{Proposition}
\newtheorem{coro}{Corollary}

\definecolor{myblue}{rgb}{0,0.1,0.9}
\definecolor{myred}{rgb}{0.9,0.1,0}

\title{Coalescent histories for lodgepole species trees}
\author{Filippo Disanto\thanks{Corresponding author. Email: fdisanto@stanford.edu.} \\
Noah A.~Rosenberg \\
\\ {\small Department of Biology, Stanford University, Stanford, CA 94305 USA} }

\begin{document}

\maketitle

\begin{abstract}
Coalescent histories are combinatorial structures that describe for a given gene tree and species tree the possible lists of branches of the species tree on which the gene tree coalescences take place. Properties of the number of coalescent histories for gene trees and species trees affect a variety of probabilistic calculations in mathematical phylogenetics. Exact and asymptotic evaluations of the number of coalescent histories, however, are known only in a limited number of cases. Here we introduce a particular family of species trees, the \emph{lodgepole} species trees $(\lambda_n)_{n\geq 0}$, in which tree $\lambda_n$ has $m=2n+1$ taxa. We determine the number of coalescent histories for the lodgepole species trees, in the case that the gene tree matches the species tree, showing that this number grows with $m!!$ in the number of taxa $m$. This computation demonstrates the existence of tree families in which the growth in the number of coalescent histories is faster than exponential. Further, it provides a substantial improvement on the lower bound for the ratio of the largest number of matching coalescent histories to the smallest number of matching coalescent histories for trees with $m$ taxa, increasing a previous bound of $(\sqrt{\pi} / 32)[(5m-12)/(4m-6)] m \sqrt{m}$ to $[ \sqrt{m-1}/(4 \sqrt{e}) ]^{m}$. We discuss the implications of our enumerative results for phylogenetic computations. 
\end{abstract}


\section{Introduction}
\label{secIntroduction}

Advances in the mathematical investigation of gene genealogies and the increasing availability of genetic data from diverse taxa have clarified that species trees, representing the branching histories of populations of organisms, need not be reflected in gene trees that represent the histories of individual genomic regions \citep{PamiloandNei88, Maddison97, Nichols01}. New developments concerning the relationship between gene trees and species trees have now led to new methods for species tree inference, new approaches to inferences about evolutionary phenomena from gene tree discordance, and an improved understanding of the branching descent process \citep{DegnanAndRosenberg09, LiuEtAl09:mpe, KnowlesandKubatko10}.

Investigations of the evolution of genomic regions along the branches of species trees have also  generated new combinatorial structures that can assist in studying gene trees and species trees \citep{Maddison97, DegnanAndSalter05, ThanAndNakhleh09, Wu12}. Among these structures are coalescent histories, structures that for a given gene tree topology and species tree topology represent possible pairings of coalescences in the gene tree with branches of the species tree on which the coalescences take place \citep{DegnanAndSalter05, Rosenberg07:jcb}.

Coalescent histories are important in a number of types of studies of the relationship between gene trees and species trees. They have appeared in empirical investigations of the gene tree topologies likely to be produced along the branches of a given species tree \citep{RosenbergAndTao08}. They are a component of mathematical proofs that concern properties of evolutionary models of gene trees conditional on species trees \citep{AllmanEtAl11:jmathbiol, ThanAndRosenberg11}. Coalescent histories also arise in studying state spaces for models that consider transitions along the genome among the gene genealogies represented at specific sites \citep{HobolthEtAl07, HobolthEtAl11, DutheilEtAl09}.

Many coalescent histories might be possible for a given gene tree and species tree, and the number of possible coalescent histories is a key quantity in the study of gene trees and species trees. In particular, because the probability of a gene tree topology conditional on a species tree can be written as a sum over coalescent histories \citep{DegnanAndSalter05}, the time required for computing gene tree probabilities is proportional to the number of coalescent histories compatible with a given gene tree and species tree. Thus, to study computational aspects of the use of coalescent histories, it has been of interest to evaluate the number of coalescent histories permissible for a given pair consisting of a gene tree and a species tree.

\cite{DegnanAndSalter05}, who initiated the study of coalescent histories, reported that if the labeled gene tree topology and species tree topology have the same matching ``caterpillar" shape with $m$ taxa, then the number of coalescent histories is the Catalan number,
\begin{equation}\label{catalano1}
c_{m-1}=\frac{1}{m}{2m-2 \choose m-1}.
\end{equation}
The Catalan sequence $c_m$ is asymptotic to $4^m/(m^{3/2} \sqrt{\pi})$. \cite{Rosenberg07:jcb} and \cite{ThanEtAl07} provided recursive procedures that list all possible coalescent histories given a gene tree and species tree, and \cite{Rosenberg07:jcb} offered simple recursive formulas for counting them. \cite{Rosenberg07:jcb, Rosenberg13:tcbb} and \cite{RosenbergAndDegnan10} then solved the recursion in a number of specific cases.


\begin{figure}[tpb]
\begin{center}
\includegraphics*[scale=0.62,trim=0 0 0 0]{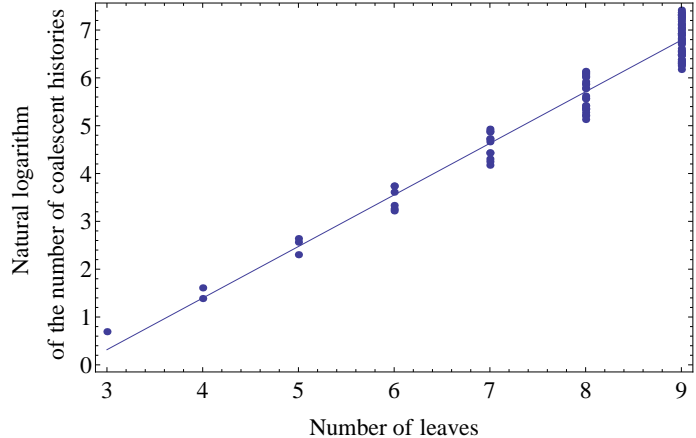}
\end{center}
\vspace{-.7cm}
\caption{{\small Natural logarithm of the number of coalescent histories for all matching gene trees and species trees with at most 9 taxa. The values plotted are taken from Tables 1-4 of \cite{Rosenberg07:jcb}. Each dot corresponds to a tree of the specified size. The line represents a linear regression $y=a+bx$, with $a \approx -2.91891$ and $b \approx 1.07865$.
}} \label{figPlot1}
\end{figure}

What is the asymptotic behavior of the number of coalescent histories as the number of taxa increases? In Figure~\ref{figPlot1}, we show values taken from \cite{Rosenberg07:jcb} for the number of coalescent histories for matching gene trees, for all species trees with $m \leq 9$ taxa. On a logarithmic scale, a linear model fits the values quite well, suggesting that in general, the number of coalescent histories for matching gene trees and species trees might grow exponentially in the number of taxa. Existing enumerations of coalescent histories in particular cases, both for the caterpillar trees in eq.~\ref{catalano1} and in related caterpillar-like families \citep{Rosenberg13:tcbb}, support this prediction. We might therefore expect that for a generic family of species trees of increasing size, the number of coalescent histories for the matching gene tree increases exponentially.


Here, we show that this prediction does not always hold. Indeed, we exhibit a family of species trees $(\lambda_n)_n$---that we term the \emph{lodgepole} family---whose number of coalescent histories grows with the double factorials, and thus increases at a rate that is faster than exponential in the number of taxa. We use the lodgepole family to further understand the variability at a given $m$ of the number of coalescent histories for cases with matching gene trees and species trees. \cite{Rosenberg07:jcb} obtained a lower bound on the ratio of the largest number of coalescent histories to the smallest number of coalescent histories at $m$ taxa, showing that this ratio was greater than a constant multiple of $(\sqrt{\pi} / 32)[(5m-12)/(4m-6)] m \sqrt{m}$. Here we improve substantially upon this lower bound, demonstrating that it exceeds the much larger $\big[ \sqrt{m-1}/(4 \sqrt{e}) \big]^{m}$.


\section{Preliminaries}
\label{secPreliminaries}

\subsection{Species trees and coalescent histories}
\label{secPreliminaries2}

A \emph{species tree} is a binary rooted tree equipped with a labeling for the leaves. As in other studies of coalescent histories, a single labeling can without loss of generality be taken as representative of an unlabeled species tree topology. When the labeling is not needed, we abbreviate the arbitrarily labeled species tree by its unlabeled shape and consider the labeled and unlabeled topologies interchangeably. We consider matching gene trees and species trees with the same labeled topology $t$.

\begin{figure}[tpb]
\begin{center}
\includegraphics*[scale=0.78,trim=0 0 0 0]{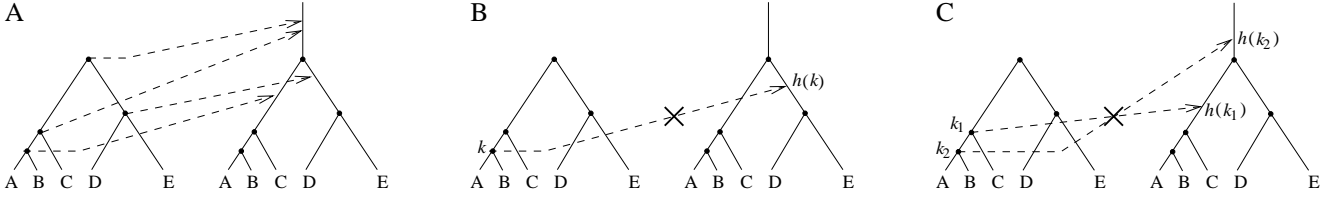}
\end{center}
\vspace{-.7cm}
\caption{{\small Coalescent histories with matching gene trees and species trees. (A) A matching coalescent history. (B) Condition (a) for matching coalescent histories is violated because leaf $B$ descends from node $k$ but not from branch $h(k)$. (C) Condition (b) for matching coalescent histories is violated, as node $k_2$ descends from node $k_1$ but the branch $h(k_2)$ remains strictly above the branch $h(k_1)$.}} \label{figPappa2}
\end{figure}

We term a coalescent history for the case when the gene tree and species tree have the same labeled topology a \emph{matching coalescent history}. Given a species tree $t$, a mapping $h$ from the internal nodes of $t$ to the branches of $t$ is a matching coalescent history of $t$ when it satisfies both of the following two conditions: (a) for all leaves $x$ in $t$, if $x$ descends from internal node $k$ in $t$, then $x$ descends from branch $h(k)$ in $t$; (b) for all internal nodes $k_1$ and $k_2$ in $t$, if $k_2$ is a descendant of $k_1$ in $t$, then branch $h(k_2)$ is descended from or coincides with branch $h(k_1)$ in $t$. Figure~\ref{figPappa2}A shows an example of a matching coalescent history. The examples in Figure~\ref{figPappa2}B and \ref{figPappa2}C are not matching coalescent histories; in Figure~\ref{figPappa2}B, condition (a) is violated, and in Figure~\ref{figPappa2}C, condition (b) is violated.

\subsection{The lodgepole family of species trees}
\label{secLodgepole}

\begin{figure}[tpb]
\begin{center}
\includegraphics*[scale=.84,trim=0 0 0 0]{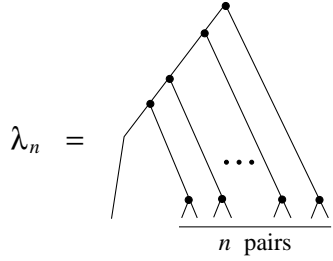}
\end{center}
\vspace{-.7cm}
\caption{{\small  The lodgepole family of species trees $\lambda_n$.
Starting from the tree with one taxon ($\lambda_0$), by adding $n \geq 0$ cherries, we obtain the tree $\lambda_n$. The term \emph{lodgepole} is after the lodgepole pine tree, \emph{Pinus contorta}, one of a number of pine species in which needles extend from the main twig in bundles of two.}} \label{figFamilies3}
\end{figure}

We focus here on the number of matching coalescent histories (\emph{histories} or \emph{coalescent histories} for short) for a particular family of species trees, $(\lambda_n)_{n\geq 0}$, that we call the \emph{lodgepole} family. We define $\lambda_0$ as the 1-taxon tree. For $n \geq 0$, we inductively define $\lambda_{n+1}$ by appending $\lambda_n$ and a tree with two leaves (a \emph{cherry}) to a common root (Fig.~\ref{figFamilies3}). Note that the tree $\lambda_n$ has $m=2n+1$ rather than $n$ taxa; we use $n$ to denote the $n$th tree $\lambda_n$ of the lodgepole family and perform our enumerations according to this parameter, later returning to $m$, the number of taxa. We view $\lambda_n$ as unlabeled, or as having an arbitrary labeling.

The lodgepole family $(\lambda_n)_{n \geq 0}$ can be seen as a modification of the caterpillar family of species trees, in which a family of trees is generated by sequentially appending the previous tree in the family and a single taxon---instead of a cherry, as in the lodgepole family---to a common root.

\subsection{Dyck paths}
\label{secDyck}

\begin{figure}[tpb]
\begin{center}
\includegraphics*[scale=.66,trim=0 0 0 0]{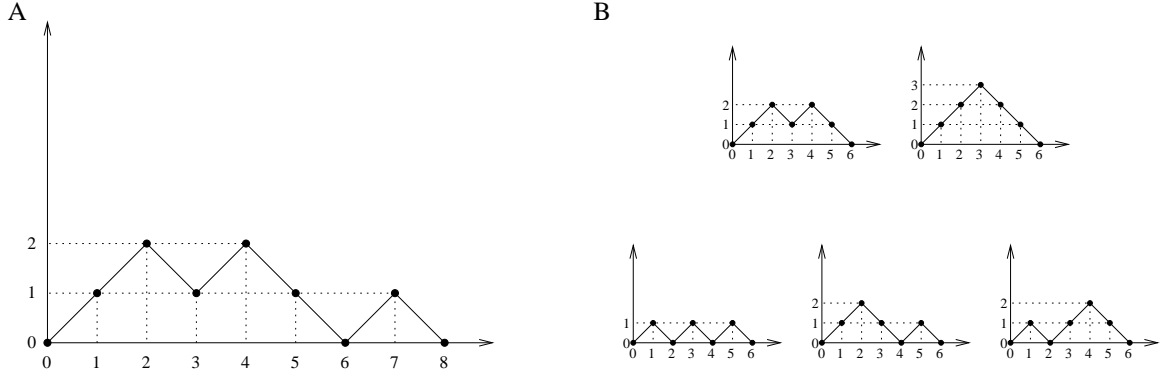}
\end{center}
\vspace{-.7cm}
\caption{{\small Dyck paths. (A) The Dyck path of size 4 whose sequence of steps is $UUDUDDUD$. It contains 4 unit up-steps $U$ and 4 unit down-steps $D$ that never pass strictly below the $x$-axis. (B) The five possible Dyck paths of size 3. The two at the top are indecomposable because they touch the $x$-axis only at the endpoints $(0,0)$ and $(6,0)$.}} \label{figDyck4}
\end{figure}

To enumerate histories for lodgepole species trees, we make use of results that involve certain lattice paths, the \emph{Dyck} paths \citep{Stanley99}. A Dyck path of size $n$ is a lattice path that starts from $(0,0)$ and ends at $(2n,0)$ in the quarter plane, that has $n$ unit steps up (each labeled $U$) and $n$ unit steps down (labeled $D$), and that never passes below the $x$-axis (Fig.~\ref{figDyck4}A). It is useful to distinguish the \emph{indecomposable} Dyck paths from the \emph{decomposable} ones. A Dyck path of size $n$ is said to be indecomposable when it touches the $x$-axis only at the extreme points $(0,0)$ and $(2n,0)$. A Dyck path is decomposable if it is not indecomposable. In Figure~\ref{figDyck4}B, the two Dyck paths at the top are indecomposable, and the remaining three are decomposable.


\section{The number of matching coalescent histories for lodgepole species trees}
\label{secNumber}

\subsection{Overview}
\label{secOverview}

We are now ready to compute the number $h_n$ of matching coalescent histories for the lodgepole tree~$\lambda_n$. We start in Section~\ref{secFirst} by obtaining a combinatorial formula that computes $h_n$ as a sum over a certain set of vectors $V_n$. In Section~\ref{secCorrespondence}, we show that by a bijection of coalescent histories for $\lambda_n$ with a certain set of Dyck paths $D_n$---a set that is in turn related to structures known as indecomposable \emph{histoires d'Hermite}---we can apply existing enumerative results to obtain a recursion for $h_n$. Finally, in Section~\ref{secAsymptotic}, we study the asymptotic behavior of $h_n$.


\subsection{A first combinatorial formula for $h_n$}
\label{secFirst}

For $n \geq 1$, we define a set $V_n$ of integer vectors $\overline{a} = (a_1,a_2,...,a_n)$ as
\begin{equation*}
V_n=\{ \overline{a} : \, a_1=2 \text{\, and \,} 2 \leq a_i \leq a_{i-1}+1 \text{\, for \,} 2\leq i \leq n \}.
\end{equation*}
Setting, for instance, $n=3$, we obtain $V_3 = \{ (2,2,2),(2,2,3),(2,3,2),(2,3,3),(2,3,4) \}$.

\begin{figure}[tpb]
\begin{center}
\includegraphics*[scale=.60,trim=0 0 0 0]{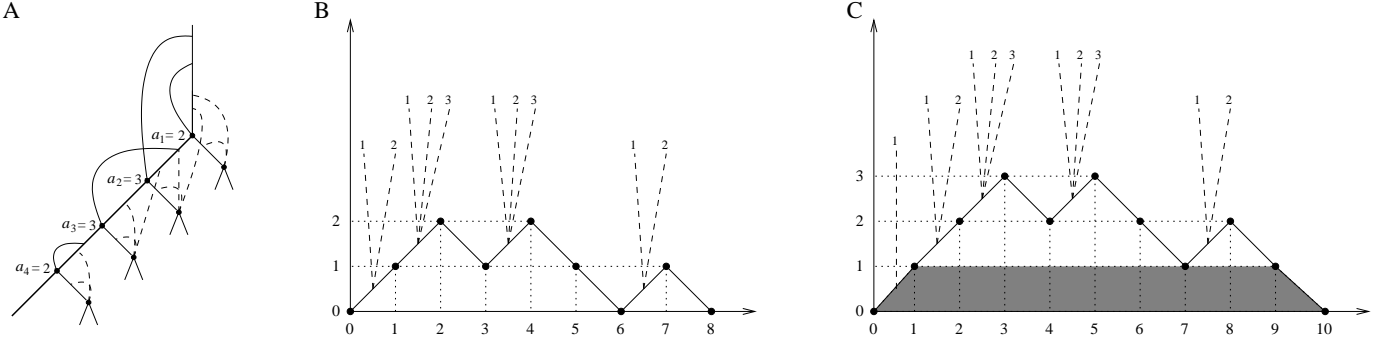}
\end{center}
\vspace{-.7cm}
\caption{{\small Combinatorial structures for computation of $h_n$: coalescent histories, labeled Dyck paths, and histoires d'Hermite.
(A) Coalescent histories of $\lambda_4$. Arcs represent the mapping of the nodes of $\lambda_4$ to its branches. Each history of $\lambda_n$ can be constructed in two steps. First, a mapping of the nodes of the main branch to branches of the tree is fixed. Next, a compatible mapping of the cherry nodes is constructed. For the nodes of the main branch, the mapping in the figure is encoded by the vector $\overline{a}=(a_1,a_2,a_3,a_4)=(2,3,3,2) \in V_4$: for $i=1,2,3,4$, the $i$th node of the main branch is mapped onto the branch $a_i-2$ nodes above it (solid arcs). Dashed arcs represent possible mappings of the cherry nodes that are  compatible with the mapping for the main branch determined by the vector $\overline{a}$. The $i$th cherry node can be mapped in exactly $a_i$ compatible ways.
(B) A labeled Dyck path of size 4, encoding the vector $(a_1,a_2,a_3,a_4)=(2,3,3,2) \in V_4$ from (A). The ordinate $y_i$ of the endpoint of the $i$th up-step $U_i$ satisfies $y_i=a_i-1$. By labeling each up-step $U_i$ of the path with an integer $\ell(U_i) \in [1, y_i+1]$, we obtain a path in $D_4$. The number of ways that the underlying Dyck path can be labeled is thus given by the product $\prod_{i=1}^{n=4} a_i = 36$.
(C) The indecomposable histoire d'Hermite of size 5 associated with the Dyck path of size 4 in (B). The histoire is obtained by adding an up-step labeled 1 at the beginning of the path in (B) and a down-step at the end and keeping the labels of the remaining up-steps as in (B). In this way, the $i$th up-step $U_i$ of the histoire in (C) has an integer label $\ell^*(U_i) \in [1,y_i]$, where, as in (B), $y_i$ is the ordinate of the endpoint of the $i$th up-step.}} \label{figMalleolo5}
\end{figure}

We have the following combinatorial formula to compute, for $n \geq 1$, the number of matching coalescent histories $h_n$ for the lodgepole species tree $\lambda_n$:
\begin{equation}
\label{ly}
h_n  = \sum_{\overline{a} \in V_n} \prod_{i=1}^{n} a_i.
\end{equation}
Eq.~\ref{ly} can be justified by formulating the procedure of \cite{Rosenberg07:jcb} for tabulating coalescent histories specifically in the lodgepole case, observing that a history of $\lambda_n$ can be constructed in two steps. In the tree $\lambda_n$, it is convenient to distinguish a main branch, that is, the one from which the $n$ cherry nodes in $\lambda_n$ descend (Fig.~\ref{figMalleolo5}A). The main branch of $\lambda_n$ thus contains $n$ internal nodes that we treat as ordered from the root (the first node) toward the single leaf at the end of the branch. In step (a), we fix a history for the nodes of the main branch, ignoring the attached cherries. In Figure \ref{figMalleolo5}A, this history is represented by the solid arcs: each arc maps a node of the main branch onto a permissible branch. In step (b), we choose a mapping for the cherry nodes (dashed arcs in the figure). The choice for the mapping of the cherry nodes must be compatible with the mapping in step (a) for the nodes of the main branch of $\lambda_n$. As required by the definition of coalescent histories in Section~\ref{secPreliminaries2}, the image of a cherry node $k$ cannot be placed on a branch above the one chosen in step (a) as the image of the node of the main branch to which node $k$ is appended.

The two-step procedure translates into eq.~\ref{ly}. In fact, each possible history of the main branch of $\lambda_n$ can be bijectively encoded by a vector of integers $(a_1,...,a_n) \in V_n$ by noting that the $i$th node of the main branch is mapped exactly $a_i-2$ nodes above it, associating each node with its immediate ancestral branch (Fig.~\ref{figMalleolo5}A). Once the vector has been fixed, the cherry node appended to the $i$th node of the main branch can be mapped in exactly $a_i$ compatible ways. Therefore, with the sum in eq.~\ref{ly}, we are considering all the possible histories of the main branch---those constructed in step (a)---and for each of these histories, the product counts the number of compatible mappings of the cherry nodes as considered in step (b).

By applying eq.~\ref{ly}, setting $h_0=1$ for convenience, we computed the first terms of the sequence $h_n$ (Table \ref{tableNumbers}). The values for $n=1,2,3,4$ accord with the values computed in the enumerations of coalescent histories reported for small trees in Tables 1 and 4 of \cite{Rosenberg07:jcb}.

\subsection{Correspondence with the histoires d'Hermite and a recursion for $h_n$}
\label{secCorrespondence}

We now show that a bijective correspondence exists between histories of the lodgepole family $(\lambda_n)_{n \geq 0}$ and certain labeled paths in the plane. Indeed, note that as in the example in Figure~\ref{figMalleolo5}B, each vector $\overline{a} \in V_n$ bijectively encodes a Dyck path of size $n \geq 1$.
More precisely, starting from the vector $\overline{a}$, a Dyck path with $n$ up-steps is uniquely determined by fixing for each $i$ the ordinate $y_i$ of the endpoint of the $i$th up-step according to
\begin{equation}
\label{giggi}
a_i = y_i+1.
\end{equation}
For instance, in Figure~\ref{figMalleolo5}B, we depict the Dyck path $UUDUDDUD$ associated with the vector $\overline{a}=(2,3,3,2) \in V_4$: in fact, as in eq.~\ref{giggi}, we have $y_1=1=a_1-1, y_2=2=a_2-1, y_3=2=a_3-1,$ and $y_4=1=a_4-1$.

Defining $D_n$ as the set of Dyck paths of size $n$ that for each $i$ have the $i$th up-step $U_i$ labeled by an integer
\begin{equation}
\label{elle}
\ell(U_i) \in [1, y_i+1],
\end{equation}
we can also interpret eq.~\ref{ly} as the formula that computes the cardinality $|D_n|$, so that
\begin{equation}
\label{hed}
h_n = |D_n|.
\end{equation}
Eq.~\ref{hed} holds for $n \geq 0$ including the case $n=0$, as we set $h_0=1$, and by counting the empty path, $D_0 = 1$. In interpreting eq.~\ref{ly} as an enumeration of labeled Dyck paths in $D_n$, the sum in eq.~\ref{ly} traverses all possible Dyck paths of size $n$ as encoded by vectors in $V_n$. For each of these Dyck paths, the product $\prod_{i=1}^{n} a_i $ computes the number of ways that the path can be labeled. By eqs.~\ref{giggi} and \ref{elle}, each label $\ell(U_i)$ has $a_i$ possible values.

This result, similar to a bijection with monotonic paths used by \cite{Degnan05} to count coalescent histories in the caterpillar case, allows us to switch from counting the histories of $\lambda_n$ to counting labeled paths in $D_n$. The correspondence in eq.~\ref{hed} can be used to obtain a recursion for $h_n$. Starting with $h_0=1$, we have for $n\geq 1$,
\begin{equation}
\label{pijo}
h_n = (2n+1)!! - \sum_{k=0}^{n-1} (2k+1)!! \, h_{n-1-k}.
\end{equation}
To prove eq.~\ref{pijo}, we make use of the relationship between the labeled Dyck paths in $D_n$ and the so-called \emph{histoires d'Hermite} of size $n+1$ (\emph{histoires} for short, Fig.~\ref{figMalleolo5}C).

\begin{table}
\begin{center}
{\caption{The number of matching coalescent histories.}\label{tableNumbers}}
\begin{tabular}{lp{1.3in}p{1.3in}p{1.3in}p{1.3in}}
\hline
Number of taxa $m$ & \multicolumn{4}{c}{Number of matching coalescent histories} \\
($m=2n+1$)         & \multicolumn{4}{c}{}                                        \\ \hline
                   & Predicted by the linear regression model
                   & Exact value of $h_{(m-1)/2}$ for the lodgepole tree
                   & Upper bound for $h_m^-$ based on ``bicaterpillars''
                   & Lower bound for $h_m^+$ based on lodgepole trees \\ \hline
3                  & 1          & 2           & 2          & 2              \\
5                  & 12         & 10          & 10         & 14             \\
7                  & 103        & 74          & 65         & 138            \\
9                  & 888        & 706         & 481        & 1,663          \\
11                 & 7,679      & 8,162       & 5,544      & 6,237          \\
13                 & 66,406     & 110,410     & 56,628     & 90,090         \\
15                 & 574,261    & 1,708,394   & 613,470    & 1,447,875      \\
17                 & 4,966,073  & 29,752,066  & 6,952,660  & 25,844,568     \\
19                 & 42,945,396 & 576,037,442 & 81,662,152 & 509,233,725    \\ \hline
\end{tabular}
\end{center}
Given $m$, we exponentiate the value from the regression model in Figure \ref{figPlot1} and round to the nearest integer. The exact $h_{(m-1)/2}=h_n$ is computed from eq.~\ref{ly} or \ref{pijo}. For $m \leq 9$, the upper bound for $h_m^-$ and the lower bound for $h_m^+$ are computed exactly from Tables 1-4 of \cite{Rosenberg07:jcb}. For $m \geq 11$, the upper bound for $h_m^-$ is computed as $c_n c_{n+1}$, with $c_n$ as in eq.~\ref{catalano1}. Applying Proposition \ref{noia} and noting that $(n-2)/n = (m-5)/(m-1)$, the lower bound for $h_m^+$ is computed as $m!! \, (m-5)/(m-1)$, rounding down where necessary.
\end{table}

An \emph{histoire} of size $n \geq 1$ is a labeled Dyck path of size $n$, but with a labeling scheme $\ell^*(U_i)$ for its up-steps that slightly differs from the scheme $\ell(U_i)$ considered in eq.~\ref{elle} for the paths of $D_n$. Indeed, in an histoire, for each $i$, the $i$th up-step $U_i$ carries an integer label
\begin{equation}
\label{ellestella}
\ell^*(U_i) \in [1,y_i],
\end{equation}
where $y_i$ is, as before, the ordinate of the endpoint of step $U_i$. Note that for histoires of size $n$, we have $y_i$ possible values for each label $\ell^*(U_i)$, whereas for Dyck paths in $D_n$, we had $y_i+1$ possibilities for label $\ell(U_i)$.

Denote by $H_n$ the set of histoires of size $n$. Section 1.2 of \cite{RobletAndViennot96} found that for $n\geq 1$,
\begin{equation}
\label{storie}
|H_n| = (2n-1)!! = (2n-1) \times (2n-3) \times \ldots 3 \times 1.
\end{equation}
We say that an histoire of size $n$ is \emph{indecomposable} if its underlying Dyck path is indecomposable
(Fig.~\ref{figMalleolo5}C). It can be observed that, denoting by $H_n^\prime$ the number of indecomposable histoires of size $n$, we have for $n \geq 0$
\begin{equation}
\label{net0}
|D_n| = |H_{n+1}^{\prime}|.
\end{equation}
Indeed, as depicted in Figure~\ref{figMalleolo5}, panels B and C, each labeled Dyck path $P \in D_n$ can be bijectively mapped onto a labeled path $P^\prime \in  H_{n+1}^{\prime}$ that is obtained by adding an up-step $U$ labeled with the integer $1$ and a down-step $D$ respectively at the beginning and at the end of $P$ and keeping unchanged the labels of the remaining up-steps of $P$. In symbols, we have for $n\geq 0$ the bijective correspondence
\begin{equation}
\label{bige}
P \in D_n \Leftrightarrow U P D = P^\prime \in H_{n+1}^{\prime}.
\end{equation}
Note in fact that according to the labeling scheme $\ell^*$ for histoires (eq.~\ref{ellestella}), in $P^\prime$ only the label $\ell^*(U_1) = 1$ is possible for the new first up-step $U_1$. The up-step $U_1$ has ordinate $1$. Furthermore, for all $i$ with $1\leq i \leq n$, the ordinate $y_{i+1}^\prime$ of the $(i+1)$th up-step in $P^\prime$ satisfies
$$y_{i+1}^\prime = y_i+1,$$
where $y_i$ is the ordinate of the $i$th up-step in $P$. Therefore, keeping the labels of the up-steps of $P$ unchanged, the labeling scheme $\ell^*$ for the histoires is satisfied by $P^\prime$, as can be seen by comparing eq.~\ref{elle} and eq.~\ref{ellestella}. Finally, by construction, the path $P^\prime$ touches the $x$-axis only in the extreme points and is by definition indecomposable. By the bijection in eq.~\ref{bige}, we thus have eq.~\ref{net0}.

\begin{figure}[tpb]
\begin{center}
\includegraphics*[scale=.90,trim=0 0 0 0]{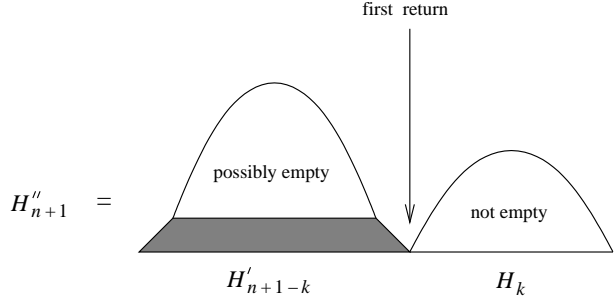}
\end{center}
\vspace{-.7cm}
\caption{{\small A decomposition of a decomposable histoire d'Hermite. Any decomposable histoire in $H_{n+1}^{\prime \prime}$ for $n \geq 1$ is uniquely obtained by concatenating an indecomposable histoire in $H_{n+1-k}^{\prime}$ with $1 \leq n+1-k \leq n$ (Fig.~\ref{figMalleolo5}C) and an histoire in $H_k$ with $1 \leq k \leq n$. The point at which they touch corresponds to the first return to the $x$-axis of the entire path. The shading indicates that the indecomposable histoire on the left begins with an up-step, ends with a down-step, and does not reach the x-axis within the shaded trapezoid.
}} \label{figCesena6}
\end{figure}

Combining eqs.~\ref{hed} and \ref{net0}, we obtain for $n \geq 0$
\begin{equation}
\label{net}
h_n = |H_{n+1}^{\prime}|.
\end{equation}
We denote by $H_n^{\prime\prime}$ the set of \emph{decomposable} (not indecomposable) histoires of size $n \geq 1$. An histoire in $H_{n+1}^{\prime\prime}$ can be decomposed uniquely as a concatenation of an indecomposable histoire in $H_{n+1-k}^\prime$ for some $k$, $1 \leq k \leq n$, and a second histoire that is either decomposable or indecomposable and hence lies in $H_k$ (Fig.~\ref{figCesena6}). The endpoint of the indecomposable histoire in $H_{n+1-k}^\prime$ provides the first return of the decomposable histoire in $H_{n+1}^{\prime\prime}$ to the x-axis, after which the histoire in $H_k$ might or might not touch the $x$-axis at a point in its interior.

Applying the decomposition, we have for $n\geq 1$
$$|H_{n+1}^{\prime \prime}| = \sum_{k=1}^{n}  |H_{n+1-k}^\prime| \, |H_{k}|.$$
Because the number of histoires in $H_n$ is known from eq.~\ref{storie}, and because each histoire is either decomposable or indecomposable, we obtain a  recursion for the number of indecomposable histoires of size $n+1$:
\begin{eqnarray}
|H^\prime_{n+1}| &=& |H_{n+1}| - |H_{n+1}^{\prime \prime}| \nonumber \\
& = & (2n+1)!! - \sum_{k=1}^{n} |H_{n+1-k}^\prime| \, |H_{k}| \nonumber \\
& = & (2n+1)!! - \sum_{k=1}^{n} |H_{n+1-k}^\prime| \, (2k-1)!!. \nonumber
\end{eqnarray}
By eq.~\ref{net}, we have demonstrated eq.~\ref{pijo}.

The fact that $h_n$ can be computed as in eq.~\ref{pijo} shows that the matching coalescent histories of $\lambda_n$ are equinumerous with other combinatorial structures. In particular, in addition to being the number of coalescent histories for lodgepole trees and the number of indecomposable histoires d'Hermite of size $n+1$, $h_n$ appears in enumerating topologically distinct Feynman diagrams of order $n$ \citep{Jacobs81, BattagliaAndGeorge88}, as well as in counting for an alphabet of size $n+1$ a class of ``irreducible'' words in which each letter appears exactly twice, and in which the first appearances of the letters appear in a canonical order \citep{BurnsAndMuche11}.


\subsection{Asymptotic behavior of $h_n$ and its consequences}
\label{secAsymptotic}

We now turn to using our recursion in eq.~\ref{pijo} to determine asymptotic properties of the number $h_n$ of matching coalescent histories for lodgepole species trees. From eq.~\ref{pijo}, it immediately follows for $n\geq 0$ that
\begin{equation}
\label{ciuccio}
h_n \leq (2n+1)!!.
\end{equation}
Therefore, dividing both sides of eq.~\ref{pijo} by (2n+1)!!, for $n \geq 1$, we can write
\begin{equation}
\label{poi}
1 \geq \frac{h_{n}}{(2n+1)!!} = 1 - \sum_{k=0}^{n-1} \frac{(2k+1)!!}{(2n+1)!!} h_{n-1-k} \geq  1 - \sum_{k=0}^{n-1} \frac{(2k+1)!! \, [2(n-1-k)+1]!!}{(2n+1)!!}.
\end{equation}
The final step in eq.~\ref{poi} follows by replacing $h_{n-1-k}$ with the upper bound $[2(n-1-k)+1]!!$ from inequality \ref{ciuccio}.

Using the fact that
\begin{equation}
\label{eqDoubleFact}
(2n+1)!! = \frac{(2n+1)!}{2^n \, n!},
\end{equation}
the sum in eq.~\ref{poi} can be simplified as
\begin{equation}
\label{esse}
s_n = \sum_{k=0}^{n-1} \frac{(2k+1)!! \, [2(n-1-k)+1]!!}{(2n+1)!!} =  \sum_{k=0}^{n-1} \frac{{{n+1}\choose{k+1}}}{{{2n+2}\choose{2k+2}}}.
\end{equation}
For $n=1$, we have $s_1=\frac{1}{3}$. In the Appendix, we show that for $n \geq 1$, the sequence $(s_n)_{n \geq 1}$ satisfies the recursion
\begin{equation}
\label{re}
(2n+3)s_{n+1} = (n+2)s_n +1.
\end{equation}


For $n \geq 1$, the upper bound
\begin{equation}
\label{essenne}
s_n \leq \frac{2}{n}
\end{equation}
can be verified by induction. Because $s_1=\frac{1}{3}$, $s_2=\frac{2}{5}$, and $s_3=\frac{13}{35}$, inequality~\ref{essenne} holds for $n=1, 2, 3$. By eq.~\ref{re}, the inductive hypothesis yields $(2n+3)s_{n+1} \leq 2(n+2)/n +1 = (3n+4)/n$, so that
$$s_{n+1} \leq \frac{3n+4}{n(2n+3)} \leq \frac{2}{n+1},$$
where the latter inequality holds for $n \geq 3.$

Therefore, from inequalities~\ref{poi} and \ref{essenne}, we have for $n \geq 1$
\begin{equation*}
1 - \frac{2}{n} \leq 1 - s_n \leq \frac{h_n}{(2n+1)!!} \leq 1,
\end{equation*}
which finally gives the bounds
\begin{equation*}
(2n + 1)!! \left( \frac{n-2}{n}  \right) \leq h_n \leq (2n + 1)!!,
\end{equation*}
and the asymptotic relationship $h_n \sim (2n+1)!!$.
We summarize our results in a proposition.
\begin{prop} \label{noia}
The number $h_n$ of matching coalescent histories for the lodgepole family $(\lambda_n)_{n \geq 0}$ is
\begin{equation}
h_n = (2n+1)!! - \sum_{k=0}^{n-1} (2k+1)!! \, h_{n-1-k},
\end{equation}
where we set $h_0 = 1$. The following bounds hold for $n\geq 1$:
\begin{equation}
\label{zucchero2}
(2n + 1)!! \left( \frac{n-2}{n}  \right) \leq h_n \leq (2n + 1)!!,
\end{equation}
and asymptotically, we have
\begin{equation}
\label{bistecca}
h_n \sim (2n+1)!! \sim  \sqrt{2} \bigg[\frac{2(n+1)}{e}\bigg]^{n+1}.
\end{equation}
\end{prop}

\medskip The asymptotic approximation in eq.~\ref{bistecca} follows from Stirling's approximation $n! \sim \sqrt{2 \pi n} (n/e)^n$ and from an equivalent form of eq.~\ref{eqDoubleFact}, $(2n+1)!!=(2n+2)!/[2^{n+1} \, (n+1)!]$. By Proposition~\ref{noia}, using an odd number $m = 2n+1$ describing the number of taxa in the lodgepole species tree $\lambda_n$, we have the following corollary.

\begin{coro}\label{pisto}
There exists a family of species trees whose number of matching coalescent histories grows faster than exponentially in the number of taxa $m$. In particular, when $m$ is odd, the number of matching coalescent histories for the lodgepole species tree $\lambda_{(m-1)/2}$ with $m \geq 1$ leaves is asymptotically
\begin{equation}
h_{(m-1)/2} \sim m!! \sim \sqrt{2} \left(\sqrt{\frac{m+1}{e}}\right)^{m+1}.
\end{equation}
\end{coro}


\medskip

\cite{Rosenberg07:jcb} studied the variability across all species trees for a fixed number of taxa $m$ of the number of matching coalescent histories by examining a ratio
$$R(m) = \frac{h_m^{+}}{h_m^{-}},$$
where $h_m^+$ denotes the number of coalescent histories for the $m$-taxon species tree with the greatest number of matching coalescent histories and $h_m^{-}$ denotes the corresponding value for the smallest number of histories ($h_m^{-}$). In Theorem~3.18, \cite{Rosenberg07:jcb} reported a lower bound on $R(m)$ for $m\geq 2$:
\begin{equation}
\label{lower1}
R(m) \geq \bigg( \frac{\sqrt{\pi}}{32} \bigg) \bigg( \frac{5m-12}{4m-6} \bigg) \, m \sqrt{m}.
\end{equation}
Our computations with lodgepole species trees substantially increase the lower bound for $h_m^+$. By using inequality~\ref{zucchero2}, we can improve on the lower bound on $R(m)$ for the case of $m$ odd.

\begin{coro}
\label{la}
Let $R(m) = h_m^{+}/h_m^{-}$ denote the ratio of the numbers of matching coalescent histories for the $m$-taxon species trees with the greatest and smallest numbers of coalescent histories. Then, for odd $m\geq7$,
\begin{equation}
\label{sciarpino2}
R(m) \geq \left( \frac{\sqrt{m-1}}{4\sqrt{e}}  \right)^{m}.
\end{equation}
\end{coro}
\emph{Proof.}
Because $m$ is odd, we fix $m=2n+1$ and switch between indexing by $m$ and by $n$. First, for $h_m^-$, \cite{Rosenberg07:jcb} considered a ``bicaterpillar'' tree from whose root descended two caterpillar subtrees, with $\lfloor m/2  \rfloor = n$ and $\lceil m/2  \rceil = n+1$ taxa. This tree has $c_{n} c_{n+1}$ histories \citep[Theorem 3.10]{Rosenberg07:jcb}, where $c_n$ is the Catalan number as in eq.~\ref{catalano1}, so that $h_m^{-} \leq c_{n} c_{n+1}$.

Now, for $h_m^+$, we use the lodgepole species tree $\lambda_n$ with $m=2n+1$ leaves to provide a lower bound on the number of matching coalescent histories for the tree with the largest number of matching coalescent histories, so that $h_m^+ \geq h_n$. By inequality \ref{zucchero2},
$$h_m^{+} \geq h_n \geq \left( 2n+1 \right)!! \left( \frac{n-2}{n} \right).$$
Therefore, with $c_n$ as in eq.~\ref{catalano1},
\begin{equation*}\label{bau}
R(m) = \frac{h_m^{+}}{h_m^{-}}  \geq  \frac{(2n+1)!!}{c_{n} c_{n+1} } \left( \frac{n-2}{n}  \right)  = \frac{(2n+1)(n-2)(n+1)(n+2)}{2^n n} \, \frac{n!}{{{2n+2}\choose{n+1}}},
\end{equation*}
where we have again used eq.~\ref{eqDoubleFact}. Using the Stirling bound $n! \geq \sqrt{2 \pi n}(n/e)^n$ and noting that ${{2n}\choose{n}} \leq 4^n$, we have
\begin{eqnarray}\nonumber
R(m) & \geq & \frac{(2n+1)(n-2)(n+1)(n+2)}{2^n n} \, \frac{\sqrt{2 \pi n} (n/e)^n}{4^{n+1}} \\\label{brutta}
     & =    & \frac{(2n+1)(n-2)(n+1)(n+2) \sqrt{2 \pi n} }{4 n} \, \bigg( \frac{n}{8e} \bigg)^n.
\end{eqnarray}
Substituting $n = (m-1)/2$ in inequality~\ref{brutta} to consider the number of taxa $m = 2n + 1$ yields
\begin{equation}\label{dentina}
R(m) \geq \frac{\sqrt{e \pi} m(m-5)(m+1)(m+3)}{4(m-1)} \left( \frac{\sqrt{m-1}}{4\sqrt{e}}  \right)^{m},
\end{equation}
which gives, if $m \geq 7$, inequality~\ref{sciarpino2} (and is in fact stronger than the simpler eq.~\ref{sciarpino2}). $\quad \Box$


\section{Discussion}
\label{secDiscussion}

We have defined the lodgepole family of species trees $(\lambda_n)_n$ and studied the growth of the number $h_n$ of matching coalescent histories for $\lambda_n$ as a function of $n$, showing that asymptotically, $h_n \sim (2n+1)!!$. For $m$ odd, the number $h_{(m-1)/2}$ of matching coalescent histories for the lodgepole species tree with $m$ taxa grows with $m!!$. Previous enumerative results for other species tree families have found that the number of coalescent histories increases only exponentially; we have demonstrated the existence of a family of species trees for which the number of matching coalescent histories grows more quickly than exponentially in the number of taxa (Corollary~\ref{pisto}).

Our results for lodgepole species trees indicate that the exponential increase in the number of matching coalescent histories observed in Figure~\ref{figPlot1} is misleading, at least in regard to the largest numbers of matching coalescent histories at a fixed number of taxa. We can consider the linear regression model obtained in Figure~\ref{figPlot1}---representing exponential growth---alongside an upper bound for $h_m^-$, the smallest number of matching coalescent histories at $m$ taxa, and our new lower bound for $h_m^+$, the largest number of matching coalescent histories at $m$ taxa (Table \ref{tableNumbers}). This comparison illustrates that whereas the linear model is reasonable at the small values of $m$ depicted in Figure \ref{figPlot1}, it becomes increasingly unreasonable in predicting $h_m^+$. Indeed, a consequence of the enumeration for lodgepole families is a substantially larger lower bound for the variability of the number of matching coalescent histories for species trees of fixed size (Corollary~\ref{la}).

The lodgepole trees differ from the caterpillars in that pairs of leaves rather than single leaves are descended from the internal nodes along the main branch. That the lodgepole species trees have such faster growth in their number of matching coalescent histories compared to the caterpillar species trees indicates that this apparently minor change in the branching structure of species trees leads to qualitatively different results in the number of histories. By contrast, it has been found that certain other changes to the caterpillars, replacing a caterpillar subtree by a non-caterpillar subtree, change the asymptotic growth in the number of matching coalescent histories only by a change to the constant multiple of the Catalan numbers, and do not change the overall growth rate~\citep{Rosenberg07:jcb, Rosenberg13:tcbb}.

The results have the implication that although the numbers of coalescent histories for relatively small species trees remain small enough for reasonable computation times involving enumerations of coalescent histories, the most challenging cases can grow more rapidly in the number of taxa than has been suggested in the cases that have been previously examined. It will be important to determine whether the challenging lodgepole scenario arises in practical settings, as well as the possibility that even more challenging families exist, for which the growth rate is even faster than in the lodgepole case.

The links in our analysis to Dyck paths and the histoires d'Hermite, and the appearance for the number of matching coalescent histories of lodgepole species trees of a sequence arising in other counting problems, identify known combinatorial structures to which coalescent histories can be related. These connections are promising for additional future computations about coalescent histories.


\section*{Acknowledgments}

We thank J.~Syring for botanical advice. We acknowledge grant support from the National Science Foundation (DBI-1146722).

\section*{Appendix}


This appendix proves eq.~\ref{re} from eq.~\ref{esse}. We define for $n \geq 1$ and $0 \leq k \leq n$, $F(k,n) = {{n+1}\choose{k+1}}/{{{2n+2}\choose{2k+2}}}$ and $R(k,n)=2n-2k+1$. We use $F(k,n)$ and $R(k,n)$ to apply the summation methods of \cite{PetkovsekEtAl96}. It can be verified algebraically that
\begin{equation}\label{efferre}
2(n+2)F(k,n) - 2(2n+3)F(k,n+1) = F(k+1,n)R(k+1,n) - F(k,n)R(k,n).
\end{equation}
Indeed, the identity follows by dividing both sides of eq.~\ref{efferre} by the nonzero $F(k,n)$ and applying the ratios
${F(k,n+1)}/{F(k,n)}=(2n-2k+1)/(2n+3)$ and ${F(k+1,n)}/{F(k,n)}= (2k+3)/(2n-2k-1).$
Summing both sides of eq.~\ref{efferre} for $k$ from $0$ to $n-1$, the right-hand side telescopes, giving a final contribution of $F(n,n)R(n,n) - F(0,n)R(0,n)$. Therefore, we obtain
$$2(n+2)\left( \sum_{k=0}^{n-1} F(k,n) \right)  -2(2n+3) \left[ \left( \sum_{k=0}^{n} F(k,n+1) \right) - F(n,n+1) \right]= F(n,n) R(n,n) - F(0,n) R(0,n).$$ Taking $s_n = \sum_{k=0}^{n-1} F(k,n)$ as in eq.~\ref{esse} yields
$$2(n+2)s_n  -2(2n+3)\left( s_{n+1} - \frac{1}{2n+3} \right) = 0, $$ from which eq.~\ref{re} immediately follows.

\bibliographystyle{chicago}
\bibliography{map3}

\begin{thebibliography}{}

\bibitem[\protect\citeauthoryear{Allman, Degnan, and Rhodes}{Allman
  et~al.}{2011}]{AllmanEtAl11:jmathbiol}
Allman, E.~S., J.~H. Degnan, and J.~A. Rhodes (2011).
\newblock Identifying the rooted species tree from the distribution of unrooted
  gene trees under the coalescent.
\newblock {\em J. Math. Biol.\/}~{\em 62}, 833--862.

\bibitem[\protect\citeauthoryear{Battaglia and George}{Battaglia and
  George}{1988}]{BattagliaAndGeorge88}
Battaglia, F. and T.~F. George (1988).
\newblock A {Pascal}-type triangle for the number of topologically distinct
  many-electron {Feynman} graphs.
\newblock {\em J. Math. Chem.\/}~{\em 2}, 241--247.

\bibitem[\protect\citeauthoryear{Burns and Muche}{Burns and
  Muche}{2011}]{BurnsAndMuche11}
Burns, J. and T.~Muche (2011).
\newblock Counting irreducible double occurrence words.
\newblock {\em Congressus Numerantium\/}~{\em 207}, 181--196.

\bibitem[\protect\citeauthoryear{Degnan}{Degnan}{2005}]{Degnan05}
Degnan, J.~H. (2005).
\newblock {\em Gene tree distributions under the coalescent process}.
\newblock Ph.\ D. thesis, University of New Mexico, Albuquerque.

\bibitem[\protect\citeauthoryear{Degnan and Rosenberg}{Degnan and
  Rosenberg}{2009}]{DegnanAndRosenberg09}
Degnan, J.~H. and N.~A. Rosenberg (2009).
\newblock Gene tree discordance, phylogenetic inference and the multispecies
  coalescent.
\newblock {\em Trends Ecol. Evol.\/}~{\em 24}, 332--340.

\bibitem[\protect\citeauthoryear{Degnan and Salter}{Degnan and
  Salter}{2005}]{DegnanAndSalter05}
Degnan, J.~H. and L.~A. Salter (2005).
\newblock Gene tree distributions under the coalescent process.
\newblock {\em Evolution\/}~{\em 59}, 24--37.

\bibitem[\protect\citeauthoryear{Dutheil, Ganapathy, Hobolth, Mailund,
  Uyenoyama, and Schierup}{Dutheil et~al.}{2009}]{DutheilEtAl09}
Dutheil, J.~Y., G.~Ganapathy, A.~Hobolth, T.~Mailund, M.~K. Uyenoyama, and
  M.~H. Schierup (2009).
\newblock Ancestral population genomics: the coalescent hidden {Markov} model
  approach.
\newblock {\em Genetics\/}~{\em 183}, 259--274.

\bibitem[\protect\citeauthoryear{Hobolth, Christensen, Mailund, and
  Schierup}{Hobolth et~al.}{2007}]{HobolthEtAl07}
Hobolth, A., O.~F. Christensen, T.~Mailund, and M.~H. Schierup (2007).
\newblock Genomic relationships and speciation times of human, chimpanzee, and
  gorilla inferred from a coalescent hidden {Markov} model.
\newblock {\em PLoS Genet.\/}~{\em 3}, 294--304.

\bibitem[\protect\citeauthoryear{Hobolth, Dutheil, Hawks, Schierup, and
  Mailund}{Hobolth et~al.}{2011}]{HobolthEtAl11}
Hobolth, A., J.~Y. Dutheil, J.~Hawks, M.~H. Schierup, and T.~Mailund (2011).
\newblock Incomplete lineage sorting patterns among human, chimpanzee, and
  orangutan suggest recent orangutan speciation and widepsread selection.
\newblock {\em Genome Res.\/}~{\em 21}, 349--356.

\bibitem[\protect\citeauthoryear{Jacobs}{Jacobs}{1981}]{Jacobs81}
Jacobs, A.~E. (1981).
\newblock Number of {Feynman} diagrams in arbitrary order of perturbation
  theory.
\newblock {\em Phys. Rev. D\/}~{\em 23}, 1760--1763.

\bibitem[\protect\citeauthoryear{Knowles and Kubatko}{Knowles and
  Kubatko}{2010}]{KnowlesandKubatko10}
Knowles, L.~L. and L.~S. Kubatko (Eds.) (2010).
\newblock {\em Estimating Species Trees}.
\newblock New York: Wiley.

\bibitem[\protect\citeauthoryear{Liu, Yu, Kubatko, Pearl, and Edwards}{Liu
  et~al.}{2009}]{LiuEtAl09:mpe}
Liu, L., L.~L. Yu, L.~Kubatko, D.~K. Pearl, and S.~V. Edwards (2009).
\newblock Coalescent methods for estimating phylogenetic trees.
\newblock {\em Mol. Phylogenet. Evol.\/}~{\em 53}, 320--328.

\bibitem[\protect\citeauthoryear{Maddison}{Maddison}{1997}]{Maddison97}
Maddison, W.~P. (1997).
\newblock Gene trees in species trees.
\newblock {\em Syst. Biol.\/}~{\em 46}, 523--536.

\bibitem[\protect\citeauthoryear{Nichols}{Nichols}{2001}]{Nichols01}
Nichols, R. (2001).
\newblock Gene trees and species trees are not the same.
\newblock {\em Trends Ecol. Evol.\/}~{\em 16}, 358--364.

\bibitem[\protect\citeauthoryear{Pamilo and Nei}{Pamilo and
  Nei}{1988}]{PamiloandNei88}
Pamilo, P. and M.~Nei (1988).
\newblock Relationships between gene trees and species trees.
\newblock {\em Mol. Biol. Evol.\/}~{\em 5}, 568--583.

\bibitem[\protect\citeauthoryear{{Petkov\v{s}ek}, Wilf, and
  Zeilberger}{{Petkov\v{s}ek} et~al.}{1996}]{PetkovsekEtAl96}
{Petkov\v{s}ek}, M., H.~S. Wilf, and D.~Zeilberger (1996).
\newblock {\em A=B}.
\newblock Wellesley, MA: Peters.

\bibitem[\protect\citeauthoryear{Roblet and Viennot}{Roblet and
  Viennot}{1996}]{RobletAndViennot96}
Roblet, E. and X.~G. Viennot (1996).
\newblock Th{\'e}orie combinatoire des {T}-fractions et approximants de
  {Pad\'e} en deux points.
\newblock {\em Discrete Math.\/}~{\em 153}, 271--288.

\bibitem[\protect\citeauthoryear{Rosenberg}{Rosenberg}{2007}]{Rosenberg07:jcb}
Rosenberg, N.~A. (2007).
\newblock Counting coalescent histories.
\newblock {\em J. Comput. Biol.\/}~{\em 14}, 360--377.

\bibitem[\protect\citeauthoryear{Rosenberg}{Rosenberg}{2013}]{Rosenberg13:tcbb}
Rosenberg, N.~A. (2013).
\newblock Coalescent histories for caterpillar-like families.
\newblock {\em IEEE/ACM Trans. Comp. Biol. Bioinf.\/}~{\em 10}, 1253--1262.

\bibitem[\protect\citeauthoryear{Rosenberg and Degnan}{Rosenberg and
  Degnan}{2010}]{RosenbergAndDegnan10}
Rosenberg, N.~A. and J.~H. Degnan (2010).
\newblock Coalescent histories for discordant gene trees and species trees.
\newblock {\em Theor. Pop. Biol.\/}~{\em 77}, 145--151.

\bibitem[\protect\citeauthoryear{Rosenberg and Tao}{Rosenberg and
  Tao}{2008}]{RosenbergAndTao08}
Rosenberg, N.~A. and R.~Tao (2008).
\newblock Discordance of species trees with their most likely gene trees: the
  case of five taxa.
\newblock {\em Syst. Biol.\/}~{\em 57}, 131--140.

\bibitem[\protect\citeauthoryear{Stanley}{Stanley}{1999}]{Stanley99}
Stanley, R.~P. (1999).
\newblock {\em Enumerative Combinatorics Volume 2}.
\newblock New York: Cambridge University Press.

\bibitem[\protect\citeauthoryear{Than and Nakhleh}{Than and
  Nakhleh}{2009}]{ThanAndNakhleh09}
Than, C. and L.~Nakhleh (2009).
\newblock Species tree inference by minimizing deep coalescences.
\newblock {\em PLoS Comp. Biol.\/}~{\em 5}, e1000501.

\bibitem[\protect\citeauthoryear{Than, Ruths, Innan, and Nakhleh}{Than
  et~al.}{2007}]{ThanEtAl07}
Than, C., D.~Ruths, H.~Innan, and L.~Nakhleh (2007).
\newblock Confounding factors in {HGT} detection: statistical error, coalescent
  effects, and multiple solutions.
\newblock {\em J. Comput. Biol.\/}~{\em 14}, 517--535.

\bibitem[\protect\citeauthoryear{Than and Rosenberg}{Than and
  Rosenberg}{2011}]{ThanAndRosenberg11}
Than, C.~V. and N.~A. Rosenberg (2011).
\newblock Consistency properties of species tree inference by minimizing deep
  coalescences.
\newblock {\em J. Comput. Biol.\/}~{\em 18}, 1--15.

\bibitem[\protect\citeauthoryear{Wu}{Wu}{2012}]{Wu12}
Wu, Y. (2012).
\newblock Coalescent-based species tree inference from gene tree topologies
  under incomplete lineage sorting by maximum likelihood.
\newblock {\em Evolution\/}~{\em 66}, 763--775.

\end{thebibliography}

\end{document}